\documentclass[aps,prl,superscriptaddress,a4paper,twocolumn]{revtex4-1}

\usepackage{lmodern}
\usepackage{dsfont, color, graphics,yfonts}
\usepackage{amsmath,amssymb}
\usepackage{graphicx,epsfig,subcaption}
\usepackage{soul,xcolor} 

\def\beq{\begin{equation}}
\def\eeq{\end{equation}}
\def\beqa{\begin{eqnarray}}
\def\eeqa{\end{eqnarray}}
\def\bra#1{\mathinner{\langle{#1}|}}
\def\ket#1{\mathinner{|{#1}\rangle}}

\def\prjct#1{\mathinner{|{#1}\rangle}\!\!\mathinner{\langle{#1}|}}

\def\eq{\begin{equation}}
\def\eeq{\end{equation}}
\def\beqa{\begin{eqnarray}}
\def\eeqa{\end{eqnarray}}
\def\prjct#1{\mathinner{|{#1}\rangle}\!\!\mathinner{\langle{#1}|}}

\begin{document}

\setstcolor{red}

\title{Bounding quantum gravity inspired decoherence \\ using atom interferometry}

\author{Ji\v{r}\'{i} Min\'{a}\v{r}}
\affiliation{School of Physics and Astronomy, University of Nottingham, Nottingham NG7 2RD, United Kingdom}
\author{Pavel Sekatski}
\affiliation{Institut f\"ur Theoretische Physik, Universit\"at Innsbruck, Technikerstra{\ss}e 21a, A-6020 Innsbruck, Austria}
\author{Nicolas Sangouard}
\affiliation{Department of Physics, University of Basel, Klingelbergstrasse 82, 4056 Basel, Switzerland}

\begin{abstract}
Hypothetical models have been proposed in which explicit collapse mechanisms prevent the superposition principle to hold at large scales. In particular, the model introduced by Ellis and co-workers [Phys. Lett. B {\bf 221}, 113 (1989)] suggests that quantum gravity might be responsible for the collapse of the wavefunction of massive objects in spatial superpositions. We here consider a recent experiment reporting on interferometry with atoms delocalized over half a meter for timescale of a second [Nature {\bf 528}, 530 (2015)] and show that the corresponding data strongly bound quantum gravity induced decoherence and rule it out in the parameter regime considered originally. 
\end{abstract}

\maketitle

\paragraph{Introduction --} 
While the structure of space-time at the Planck scale is not known in detail, one expects that it is a subject to quantum fluctuations. It has been suggested that the departure from flat space-time on short distances can degrade the spatial coherences of massive systems. In particular, Ellis and co-workers have proposed that non-trivial configurations of space-time  -- wormholes -- might lead to the decoherence of spatial superpositions of massive objects, see the model presented in Ref. \cite{Ellis84}, and further elaborated in Ref. \cite{Ellis89}. The basic idea is that a massive object initially prepared in a superposition of two different locations entangles with degrees of freedom of the wormholes which would leave the object in a classical mixture once the wormholes are traced out. While the wormhole theory has been proposed \cite{Coleman88} and subsequently rejected as a solution to the cosmological constant problem \cite{Fischler89, Padilla15}, the model of Ellis and co-workers provides an explicit mechanism for a quantum-classical crossover. It is also an example of a phenomenological approach in which falsifiable predictions might provide hints about quantum gravity, see e.g \cite{Ellis92,Mavromatos05} for string theory inspired models (or \cite{Hossenfelder11} for a review of various phenomenological models). \\

Many proposals have been made for testing quantum gravity induced decoherence as suggested in Ref. \cite{Ellis89} including cavity quantum optomechanics with nano- and micro-mechanical oscillators \cite{Pepper12, Sekatski14, Ghobadi14, Bahrami14, Ho16} or levitating dielectric nano-spheres \cite{Romero-Isart11, Bera15}. While challenging experiments are being set up to implement these proposals, some of them forcing experimentalists to envision experiments in space \cite{Kaltenbaek16}, we have pointed out in \cite{Minar16} that simple techniques with single atoms trapped in optical lattices could be used to test efficiently the model of Ref. \cite{Ellis89}. In particular, we have shown that quantum gravity induced decoherence can have a significant effect on the state of a single atom if the latter is delocalized on centimeter scales for a time of the order of a few seconds. With this in mind, we welcome the experimental results presented recently in Ref. \cite{Kovachy15} where a quantum superposition of rubidium atoms at the half-meter scale was achieved using light-pulse interferometry. Here, we show that the corresponding data set\st{s} strong bounds on the model of Ref. \cite{Ellis89}, and rules it out in the parameter regime that has been considered originally. \\

\paragraph{Principle --} In this section, we summarize the principle of quantum gravity induced decoherence, arising from a phenomenological treatment \cite{Ellis84, Ellis89} of the interaction of matter with a dilute-gas wormhole background. We consider a system with mass $m,$ described initially by the state $\rho_0.$ In the model of Ref. \cite{Ellis89}, an unperturbed wormhole is described by a state $\rho_w = \ket{\psi_w}\bra{\psi_w}$ which is assumed to be pure and have a Gaussian wavefunction in the momentum space $\ket{\psi_w} = \int {\rm d}^3 {\mathbf p} \psi({\mathbf p}) \ket{{\mathbf p}}$, where $\psi({\bf p}) \propto e^{-p^2/(2 \sigma)^2}$. It has zero mean momentum and a spread $\sigma \sim c m_0^2/(\hbar m_{{\rm Pl}}) \sim 10^{-3}$m$ ^{-1}$ where $m_0$ is the mass of the nucleon, $m_{{\rm Pl}}$ the Planck mass, $c$ the speed of light and $\hbar$ the reduced Planck constant. 
The interaction between the wormhole and the system is treated in the input-output S-matrix formalism and is assumed to result in the  elastic  scattering  of the  wormholes on  the  system. Concretely, at any time there is a small probability amplitude that a wormhole scatters according to $\ket{{\bf p}} \to i \int d^3{\bf p}' e^{i ({\bf p}'- {\bf p}) \cdot{\bf X}} \delta(|{\bf p}|-|{\bf p}'|) \frac{F({\bf p}')}{|{\bf p'}|}\ket{{\bf p}'}$, which entangles it with system's position given by the operator ${\bf X}$. Here, $F({\mathbf p})$ are the dimensionless scattering amplitudes. Accordingly, once the wormhole is traced out, the effect of such a scattering event on the system is a decay of coherence terms $\ket{{\bf x}'}\! \bra{\bf{x}} \to r \ket{{\bf x}'}\! \bra{\bf{x}} $ by some factor $r$ which depends on the spatial spread $\mathbf x-\mathbf x'$. As long as the typical wavelength of wormholes $\frac{1}{\sigma}\gg |{\bf x}-{\bf x}'|$ dominates the system spatial spread, the decay grows quadratically with the spatial spread $r \propto |{\bf x}-{\bf x}'|^2$ . This dependence follows from the Taylor expansion of $e^{i ({\bf p}'- {\bf p}) \cdot{\bf X}}$ \cite{Ellis89}. It can be intuitively understood as $r$ is given by the overlap of two scattered modes centered at {\bf x} and {\bf x}' which carry the information about the system position \cite{footnote1} (see also \cite{Joos85,Gallis90} for a related discussion). Since such scattering events have a low probability and happen randomly, the interaction of the system with a dilute-gas of wormholes results in the addition of a time-independent localization term to the master equation 
\begin{equation}
\dot{\rho} = i [\rho, H_0] - \frac{1}{2} \int d^3\, {\bf x} d^3{\bf x}' \gamma_{\text{QG}} |{\bf x}-{\bf x}'|^2 \prjct{\bf x} \rho \prjct{\bf x'},
\label{eq:rho mod}
\end{equation}
where $H_0$ stands for the unitary evolution of the system and it is assumed that the characteriscic interaction time of a wormhole with the system is much smaller then the time scales of $H_0$. The resulting localization rate coefficient $\gamma_{\rm QG}$ depends on the wormhole momentum spread $\sigma$, the scattering amplitudes $F({\mathbf p})$ which are functions of the system mass $m$, and the density of wormholes which is assumed to be $O(1)$ per Planck volume by Ellis and co-workers and it is given by \cite{Ellis89} $\gamma_{\text{QG}} = \frac{(cm_0)^4 m^2}{(\hbar m_{{\rm Pl}})^3}$. While the modification eq. (\ref{eq:rho mod}) is appealing as an attempt to explain the absence of coherence of macroscopic objects, we argue in the next section that the experimental results of Ref. \cite{Kovachy15} are in contradiction with the model of Ref. \cite{Ellis89}, at least for the parameter regime presented in Ref. \cite{Ellis89}.\\

\paragraph{Experimental description --} Here, we briefly describe the experiment reported in Ref. \cite{Kovachy15}. The authors start by launching a Bose-Einstein condensate made with about $10^5$ $^{87}$Rb atoms in a $10$m high atomic fountain. Once launched, a sequence of pulses is applied to control the atom momenta so that the wave packet of each atom is split and recombined coherently to form the analogue of a Mach-Zehnder interferometer. In the experiment presented in Ref. \cite{Kovachy15}, the atomic wave packets get separated during a drift time $T=1.04$s after which they reach their maximum separation $d_{\rm max}$. They are then recombined to spatially overlap after another drift interval $T=1.04$s. The spatial separation $d_{\max}=n (\hbar k/m) T$ depends on the laser pulse wave number $k$, the number of photon recoils at the first atom splitter $n \hbar k$ and the atomic mass $m.$ In Ref. \cite{Kovachy15}, an atom splitter transferring up to $90 \hbar k$ photon recoils is obtained using 2$\hbar k$ Bragg transitions \cite{Chiow11}, which results in a distance $d_{\max} = 54$cm. The contrast of the interference is determined by measuring the variation of the normalized number of atoms in one of the two outputs of the interferometer. For $d_{\max} = 54$cm, the measured contrast is of $28\%.$ Here, the loss of coherence is in agreement with the measured atom loss in the interferometer (see Fig. 4b in \cite{Kovachy15}).\\

\paragraph{Bounding quantum gravity induced decoherence --} Let $d(t)$ be the time dependent function describing the atom delocalization. According to eq. (\ref{eq:rho mod}), the coherence term of a single atom delocalized over $d(t)$ evolves as
$
\langle {\mathbf x} | \rho(t) |{\mathbf x}'\rangle = e^{-\gamma_{\text{QG}} \int_0^{2T} d^2(t) dt} \langle {\mathbf x} | \rho_0 |{\mathbf x}'\rangle.
$
Assuming 
$
d(t)=d_{\max} + \frac{d_{\max}}{T}\left(\frac{T}{2\pi}\sin\frac{2\pi |t-T|}{T}-|t-T|\right),$ we have 
$
\langle {\mathbf x} | \rho(t) |{\mathbf x}'\rangle \approx e^{-0.8\,\gamma_{\text{QG}} d_{\max}^2 T} \langle {\mathbf x} | \rho_0 |{\mathbf x}'\rangle.
$
Considering a maximal initial coherence $\langle {\mathbf x} | \rho_0 |{\mathbf x}'\rangle=1/2$, the contrast of the corresponding interference
\begin{equation}
C  \approx  2 \langle {\mathbf x} | \rho_0 |{\mathbf x}'\rangle e^{-0.8\,\gamma_{\text{QG}} d_{\max}^2 T},
\end{equation}
is of order $10^{-11}$ when a single $^{87}$Rb atom is delocalized over $d_{\max} = 54$cm with a drift time of $T=1.04$s. Even if the trajectory of the atoms can be  different from the one that we have considered, realistic deviations cannot account for such large discrepancy. This suggests that either the model of Ref. \cite{Ellis89} has to be ruled out or the parameter regime originally considered has to be revisited, e.g. by assuming much lower wormhole densities. \\

\paragraph{Conclusion --} Ellis and co-workers have suggested more than 20 years ago that a hypothetical space-time configuration can degrade the coherence of spatial superposition states. We have shown that recently published data from atom interferometry rule out this model in the parameter regime originally proposed. The fact that the quantum gravity inspired model \cite{Ellis89} can be ruled out by an interference experiment with $^{87}$Rb atoms -- rather light objects -- might seem surprising at first sight. This can be understood, however, from the localization rate of the model of Ref. \cite{Ellis89} which depends quadratically on the mass of the system $m$, but also on the spatial separation $d$ between the two arms of the interferometer. Remarkably, the $d^2$ dependence of the localization rate is not unique to the model of Ref. \cite{Ellis89} and holds whenever the separation $d\ll \lambda$ is much smaller than the characteristic wavelength of the hypothetical field which induces the decoherence, be it wormholes, dark matter \cite{Riedel13,Riedel15} or dark energy, such as chameleon fields \cite{Burrage15,Hamilton15,Lemmel15}. This makes atom interferometry a very promising technique to probe these models. As future work, it might be interesting to see how the results of Ref. \cite{Kovachy15} bound other collapse models \cite{Bassi13} in the spirit of what has been done in Refs. \cite{Vinante16, Toros16} for example.\\

\paragraph{Acknowledgments --} J.M. would like to thank E. Copeland, C. Burrage and A. Padilla for useful discussions. We thank M. Ho for a careful reading of the manuscript. This work was supported by the grant EU-FET HAIRS 612862, the Swiss National Science Foundation grant number PP00P2-150579 and the Austrian Science Fund (FWF: P24273-N16, P28000-N27).


\begin{thebibliography}{1}
\bibitem{footnote1} We note that the same argumentation implies that in the other extreme $\frac{1}{\sigma}\ll |{\mathbf x}-{\mathbf x'}|$ the localization rate resulting from scattering becomes independent of the spread and saturates to a constant value.
\end{thebibliography}
\end{document}